\documentstyle[aps]{revtex}

\begin{document}
\draft
\title{Solvable Kinetic\ Gaussian\ Model\ in\ External\ Field}
\author{Jian-Yang Zhu$^{1,2,3}$\thanks{%
Electronic address: zhujy@bnu.edu.cn} and Z. R. Yang$^{1,2}$}
\address{$^1$CCAST (World Laboratory), Box 8730, Beijing 100080, China\\
$^2$Department of Physics and Institute of Theoretical Physics, Beijing
Normal University, Beijing 100875, China\thanks{%
Mailing address}\\
$^3$Department of Physics, Jiangxi Normal University, Nanchang 330027, China }
\maketitle

\begin{abstract}
In this paper, the single-spin transition dynamics is used to investigate
the kinetic Gaussian model in a periodic external field. We first derive the
fundamental dynamic equations, and then treat an isotropic $d$-dimensional
hypercubic lattice Gaussian spin system with Fourier's transformation
method. We obtain exactly the local magnetization and the equal-time pair
correlation function. The critical characteristics of the dynamical
relaxation $\tau _{{\bf q}}$, the complex susceptibility $\chi (\omega ,{\bf %
q})$, and the dynamical response are discussed. The results show that the
time evolution of the dynamical quantities and the dynamical responses of
the system strongly depend on the frequency and the wave vector of the
external field.
\end{abstract}


\section{Introduction}

The purpose of this work is to investigate the dynamical behavior of a
cooperative spin system, particularly, the dynamical critical behavior. It
is well known that a great progress in the understanding of critical
dynamics has been made, since Glauber\cite{Glauber} and Kawasaki\cite
{Kawasaki} completed their pioneering work on the time-dependent Ising
model. In the past two decades much research has been devoted to a better
understanding of dynamical behavior of various systems, and many theoretical
methods have been applied and developed\cite
{1,2,3,4,5,6,7,8,9,10,11,12,13,14,15,16,17}. Among these studies, the main
attention has been focused on the discrete spin systems, such as the Ising
model and the Potts model, and a little on the continuous symmetry $O(n)$
spin systems. Nevertheless, as far as our knowledge goes, only a few
analytical results were presented. In the present work and the previous paper%
\cite{Zhu}, we are focusing on obtaining the exact analytical results. This
is our main motivation.

The Gaussian model is a variation of the Ising model. It is a uniaxial
continuous spin model that shows different static critical behavior from the
Ising model. Although its static critical properties have been investigated
clearly, little attention has been paid to dynamical critical behavior. This
is also the reason we study the kinetic Gaussian model. Within the framework
of Glauber dynamics in our previous paper\cite{Zhu}, we have obtained
dynamical critical exponent $z=1/\nu =2$ at the critical point $K_c=b/2d$
based on rigorous analytical derivation.

To our knowledge, only the kinetic Ising model with time-dependent external
field has been investigated in detail\cite{Sides,Kubo}. The present work is
attempting to investigate further the dynamic behavior of the kinetic
Gaussian model with time-dependent external field. This paper is organized
as follows: In Sec. \ref{Sec.2}, we first summarize the basic theory of the
single-spin transition critical dynamics, and then derive the fundamental
equations of the kinetic Gaussian model in a periodic external field. In
Sec. \ref{Sec.3}, an isotropic $d$-dimensional hypercubic lattice Gaussian
spin model is treated by Fourier's transformation. We exactly obtain the
local magnetization and the equal-time spin-pair-correlation function. The
critical characteristics of the dynamical relaxation $\tau _{{\bf q}}$, the
complex susceptibility $\chi (\omega ,{\bf q})$, and the dynamical response
of the system to the time-dependent external field are investigated.
Finally, we end up the paper with concluding remarks in Sec. \ref{Sec.4}.

\section{Fundamental Equations}

\label{Sec.2}

\subsection{Basic theory of the single-spin transition critical dynamics}

A single-spin transition critical dynamics based on Glauber's theory\cite
{Glauber}, applying to both discrete-spin and continuous-spin systems, was
presented in our previous paper\cite{Zhu}. For the sake of application here
we only give a summary.

Spins system with Hamiltonian, ${\cal H}\left( \left\{ \sigma _i\right\}
\right) ,$ where $\sigma _i$ is the spin of site $i$ and can take discrete
values or continuous values, interacts with a large heat bath with
temperature $T$. The heat bath gives rise to spontaneous transition of spins
via exchange of the energy. The probability of transition of the $i$th spin
per unit time from one value $\sigma _i$ to another possible value $\hat{%
\sigma}_i$ is denoted by $W_i(\sigma _i\rightarrow \hat{\sigma}_i)$. Under
the assumption of single-spin transition, the probability distribution
function $P(\{\sigma _i\},t)$ of the system, being in the configuration $%
(\sigma _1,\sigma _2,\cdots ,\sigma _N)$ at time $t$, is governed by the
master equation

\begin{equation}
\frac d{dt}P(\{\sigma _j\},t)=\sum_i\sum_{\hat{\sigma}_i}\left[ -W_i(\sigma
_i\rightarrow \hat{\sigma}_i)P(\{\sigma _j\},t)+W_i(\hat{\sigma}%
_i\rightarrow \sigma _i)P(\{\sigma _{j\neq i}\},\hat{\sigma}_i,t)\right] ,
\label{1}
\end{equation}
where the spin transition probability satisfies the following restrictive
conditions:

(1) Ergodicity: 
\begin{equation}
\forall \sigma _j,\hat{\sigma}_j:\ W_j(\sigma _j\rightarrow \hat{\sigma}%
_j)\neq 0;  \label{2a}
\end{equation}

(2) Positivity: 
\begin{equation}
\forall \sigma _j,\hat{\sigma}_j:\ W_j(\sigma _j\rightarrow \hat{\sigma}%
_j)\geq 0;  \label{2b}
\end{equation}

(3) Normalization: 
\begin{equation}
\forall \sigma _j:\ \sum_{\hat{\sigma}_j}W_j(\sigma _j\rightarrow \hat{\sigma%
}_j)=1;  \label{2c}
\end{equation}

(4) Detailed balance: 
\begin{equation}
\forall \sigma _j,\hat{\sigma}_j:\ \frac{W_j(\sigma _j\rightarrow \hat{\sigma%
}_j)}{W_j(\hat{\sigma}_j\rightarrow \sigma _j)}=\frac{P_{{\normalsize eq}%
}(\sigma _1,...,\hat{\sigma}_j,...,\sigma _N)}{P_{{\normalsize eq}}(\sigma
_1,...,\sigma _j,...,\sigma _N)},  \label{2d}
\end{equation}
in which 
\begin{equation}
P_{{\normalsize eq}}\left( \{\sigma \}\right) =\frac 1Z\exp [-\beta {\cal H}%
(\{\sigma \})],\quad Z=\sum_{\{\sigma \}}\exp [-\beta {\cal H}(\{\sigma \})],
\label{3}
\end{equation}
where $P_{eq}$ is the equilibrium Boltzmann distribution function, $Z$ the
partition function and ${\cal H}(\{\sigma \})$ the Hamiltonian of the system.

By use of both the master Eq. (\ref{1}) and the normalized condition (\ref
{2c}), the time-evolving equations of the local magnetization and the
equal-time spin-pair-correlation function can be expressed as, respectively 
\begin{equation}
\frac d{dt}\left\langle \sigma _k(t)\right\rangle =-\left\langle \sigma
_k(t)\right\rangle +\sum_{\{\sigma _j\}}\left( \sum_{\hat{\sigma}_k}\hat{%
\sigma}_kW_k\left( \sigma _k\rightarrow \hat{\sigma}_k\right) \right)
P(\{\sigma _j\},t),  \label{4}
\end{equation}
\begin{eqnarray}
\frac d{dt}\left\langle \sigma _k(t)\sigma _l(t)\right\rangle
&=&-2\left\langle \sigma _k(t)\sigma _l(t)\right\rangle +\sum_{\{\sigma
_j\}}\left[ \sigma _k(t)\left( \sum_{\hat{\sigma}_l}\hat{\sigma}_lW_l\left(
\sigma _l\rightarrow \hat{\sigma}_l\right) \right) \right.  \nonumber \\
&&\left. +\sigma _l(t)\left( \sum_{\hat{\sigma}_k}\hat{\sigma}_kW_k\left(
\sigma _k\rightarrow \hat{\sigma}_k\right) \right) \right] P\left( \{\sigma
_j\},t\right) ,  \label{5}
\end{eqnarray}
where 
\begin{equation}
\left\langle \sigma _k(t)\right\rangle =\sum_{\{\sigma _l\}}\sigma
_k(t)P\left( \{\sigma _l\},t\right) ,  \label{6}
\end{equation}
\begin{equation}
\left\langle \sigma _k(t)\sigma _l(t)\right\rangle =\sum_{\{\sigma
_l\}}\sigma _k(t)\sigma _l(t)P\left( \{\sigma _l\},t\right) .  \label{7}
\end{equation}

Although the evolution starts with a certain initial state at $t=0$, the
system must be relaxed toward the final objective, which is an equilibrium
state characterized by $P_{eq}=(1/Z)\exp \left[ -\beta {\cal H}\left(
\left\{ \sigma \right\} \right) \right] $ in the absence of time-dependent
external field, via interaction with heat bath. In addition, it is usually
considered that the transition probabilities of the individual spins depend
merely on the momentary values of the neighboring spins as well as the
influence of the heat bath. So, even if the transition probability cannot be
derived exactly by means of microscope, the following form: 
\begin{equation}
W_i(\sigma _i\rightarrow \hat{\sigma}_i)=\frac 1{Q_i}\exp \left[ -\beta 
{\cal H}_i\left( \hat{\sigma}_i,\sum_{\left\langle ij\right\rangle }\sigma
_j\right) \right] ,  \label{8}
\end{equation}
is well chosen, where $Q_i$ is the coefficient determined by the normalized
condition (\ref{2c}). Equation (\ref{8}) means that the transition
probability from $\sigma _j$ to $\hat{\sigma}_j$ only depends on the heat
Boltzmann factor of the neighboring spins. If the system is in a periodic
low-frequency external field, Eq. (\ref{8}) is still a possible choice.

\subsection{The model and the fundamental equations}

To study further the dynamical behavior of the Gaussian spins system near
the critical point, we put the system in a periodic low-frequency external
field which may be regarded as electromagnetic wave\cite{Kubo}. The reduced
Hamiltonian of the system under consideration is

\begin{equation}
-\beta {\cal H}=\sum_{\left\langle i,j\right\rangle }K_{ij}\sigma _i\sigma
_j+\sum_ih_i\left( t\right) \sigma _i,  \label{9}
\end{equation}
where 
\begin{equation}
\beta =\frac 1{k_\beta T},\quad K_{ij}=\frac{J_{ij}}{k_\beta T},\quad
h_i\left( t\right) =\frac{H_i(t)}{k_\beta T}=\frac{H_0}{k_\beta T}\exp
\left( i\omega t-iqx_i\right) ,  \label{10}
\end{equation}
the first sum goes over all nearest-neighbor pairs of lattice and the second
over all sites. Unlike Ising spin model, the Gaussian model have two
extensions\cite{Berlin}: first, the spin can take any real value in the
range of ($-\infty ,+\infty $); second, to prevent all spins from tending to
infinity, the probability of finding a given spin between $\sigma _i$ and $%
\sigma _i+d\sigma _i$ is assumed to be the Gaussian-type distribution 
\begin{equation}
f(\sigma _i)d\sigma _i=\sqrt{\frac b{2\pi }}\exp \left[ -\frac b2\sigma
_i^2\right] d\sigma _i,  \label{11}
\end{equation}
where $b$ is a distribution constant independent of temperature.

In terms of those mentioned above, we can derive the fundamental equations
of the kinetic Gaussian model in the external field. Following Eq. (\ref{8}%
), we choose the spin-transition probability as

\begin{equation}
W_i(\sigma _i\rightarrow \hat{\sigma}_i)=\frac 1{Q_i}\exp \left[
\sum_wK_{i,i+w}\hat{\sigma}_i\sigma _{i+w}+h_i\hat{\sigma}_i\right] =\frac 1{%
Q_i}\exp \left[ E_i\hat{\sigma}_i\right] ,  \label{12}
\end{equation}
where 
\begin{equation}
E_i=\sum_wK_{i,i+w}\sigma _{i+w}+h_i,  \label{13}
\end{equation}
and $\sum_w$ means summation over nearest neighbors. As the spin variable
takes continuous values, the summation for spin turns into the integration 
\begin{equation}
\sum_\sigma \rightarrow \int_{-\infty }^\infty f(\sigma )d\sigma ,
\label{14}
\end{equation}
then the normalized factor $Q_i$ can be determined as 
\begin{equation}
Q_i=\sum_{\hat{\sigma}_i}\exp \left( E_i\hat{\sigma}_i\right) =\int \exp
\left( E_i\hat{\sigma}_i\right) f\left( \hat{\sigma}_i\right) d\hat{\sigma}%
_i=\exp \left( \frac{E_i^2}{2b}\right) ,  \label{15}
\end{equation}
and another useful combination formula can also be written as 
\begin{equation}
\sum_{\hat{\sigma}_i}\hat{\sigma}_iW_i\left( \sigma _i\rightarrow \hat{\sigma%
}_i\right) =\int \hat{\sigma}_iW_i\left( \sigma _i\rightarrow \hat{\sigma}%
_i\right) f\left( \hat{\sigma}_i\right) d\hat{\sigma}_i=E_i/b.  \label{16}
\end{equation}
Substituting Eq. (\ref{16}) into the time-evolving Eqs. (\ref{4}), (\ref{5}%
), one gets 
\begin{equation}
\frac d{dt}\langle \sigma _i(t)\rangle =-\langle \sigma _i(t)\rangle +\frac 1%
b\sum_wK_{i,i+w}\langle \sigma _{i+w}(t)\rangle +\frac 1bh_i\left( t\right) ,
\label{17}
\end{equation}
\begin{eqnarray}
\frac d{dt}\langle \sigma _i(t)\sigma _j(t)\rangle &=&-2\langle \sigma
_i(t)\sigma _j(t)\rangle +\frac 1b\sum_w\left[ K_{j,j+w}\langle \sigma
_i(t)\sigma _{j+w}(t)\rangle +K_{i,i+w}\langle \sigma _{i+w}(t)\sigma
_j(t)\rangle \right]  \nonumber \\
&&+\frac 1b\left[ h_i\left( t\right) \langle \sigma _j(t)\rangle +h_j\left(
t\right) \langle \sigma _i(t)\rangle \right] .  \label{18}
\end{eqnarray}
Equations (\ref{17}) and (\ref{18}) are the fundamental equations of the
kinetic Gaussian model in the external field.

\section{Exact Solution}

\label{Sec.3}For a $d$-dimensional isotropic hypercubic lattice, the dynamic
equations of the local magnetization (\ref{17}) and the spin-pair
correlation function (\ref{18}) can be rewritten as 
\begin{equation}
\frac d{dt}\langle \sigma _{{\bf i}}(t)\rangle =-\langle \sigma _{{\bf i}%
}(t)\rangle +\frac 1b\sum_{{\bf l}}K({\bf r}_{{\bf i}}-{\bf r}_{{\bf l}%
})\langle \sigma _{{\bf l}}(t)\rangle +\frac{\beta H_0}be^{i\omega t-i{\bf %
q\cdot r}_{{\bf i}}},  \label{19}
\end{equation}
\begin{eqnarray}
\frac d{dt}\langle \sigma _{{\bf i}}(t)\sigma _{{\bf j}}(t)\rangle
&=&-2\langle \sigma _{{\bf i}}(t)\sigma _{{\bf j}}(t)\rangle  \nonumber \\
&&+\frac 1b\left[ \sum_{{\bf l}}K({\bf r}_{{\bf i}}-{\bf r}_{{\bf l}%
})\langle \sigma _{{\bf l}}(t)\sigma _{{\bf j}}(t)\rangle +\sum_{{\bf l}}K(%
{\bf r}_{{\bf l}}-{\bf r}_{{\bf j}})\langle \sigma _{{\bf i}}(t)\sigma _{%
{\bf l}}(t)\rangle \right]  \nonumber \\
&&+\frac{\beta H_0}b\left[ \langle \sigma _{{\bf i}}(t)\rangle e^{i\omega t-i%
{\bf q\cdot r}_{{\bf j}}}+\langle \sigma _{{\bf j}}(t)\rangle e^{i\omega t-i%
{\bf q\cdot r}_{{\bf i}}}\right] ,  \label{20}
\end{eqnarray}
where 
\begin{equation}
{\bf q=}(q_{1,}q_2,\ldots ,q_d),{\bf r=}(x_1,x_2,\ldots ,x_d),{\bf i=(}%
i_1,i_2,\ldots ,i_d{\bf ),j=(}j_1,j_2,\ldots ,j_d{\bf ),\ldots ,}  \label{21}
\end{equation}
and 
\begin{equation}
K({\bf r}_{{\bf i}}-{\bf r}_{{\bf j}})=\left\{ 
\begin{array}{ll}
K, & \mbox{nearest-neighbor-pair,} \\ 
0, & \mbox{others.}
\end{array}
\right.  \label{22}
\end{equation}

Introducing Fourier's transformation 
\begin{equation}
\left\langle M({\bf q}^{\prime },t)\right\rangle =\sum_{{\bf i}}\langle
\sigma _{{\bf i}}(t)\rangle e^{i{\bf q}^{\prime }{\bf .r}_{{\bf i}}},
\label{23}
\end{equation}
\begin{equation}
\left\langle G({\bf q}^{\prime },{\bf q}^{\prime \prime },t)\right\rangle
=\sum_{{\bf k,l}}\left\langle \sigma _{{\bf k}}\left( t\right) \sigma _{{\bf %
l}}\left( t\right) \right\rangle e^{i{\bf q}^{\prime }.{\bf r}_{{\bf k}}+i%
{\bf q}^{\prime \prime }{\bf .r}_{{\bf l}}},  \label{24}
\end{equation}
which satisfy 
\begin{equation}
\frac 1N\sum_{{\bf l}}e^{i({\bf q}-{\bf q}^{\prime }).{\bf r}_{{\bf l}%
}}=\delta _{{\bf q},{\bf q}^{\prime }},  \label{25}
\end{equation}
\begin{equation}
\frac 1N\sum_{{\bf q}}e^{i{\bf q.}({\bf r}_{{\bf l}}-{\bf r}_{{\bf k}%
})}=\delta _{{\bf l,k}},  \label{26}
\end{equation}
we have 
\begin{equation}
\frac d{dt}\left\langle M({\bf q}^{\prime },t)\right\rangle +\left[ 1-\frac 1%
bK({\bf q}^{\prime }{\bf )}\right] \left\langle M({\bf q}^{\prime
},t)\right\rangle =\frac{NH_0\beta }b\delta _{{\bf q,q}^{\prime }}e^{i\omega
t},  \label{27}
\end{equation}
\begin{eqnarray}
&&\frac d{dt}\left\langle G({\bf q}^{\prime },{\bf q}^{\prime \prime
},t)\right\rangle +\left[ \left( 1-\frac 1bK({\bf q}^{\prime }{\bf )}\right)
+\left( 1-\frac 1bK({\bf q}^{\prime \prime }{\bf )}\right) \right]
\left\langle G({\bf q}^{\prime },{\bf q}^{\prime \prime },t)\right\rangle 
\nonumber \\
&=&\frac{NH_0\beta }b\left[ \left\langle M({\bf q}^{\prime },t)\right\rangle
\delta _{{\bf q,q}^{\prime \prime }}+\left\langle M({\bf q}^{\prime \prime
},t)\right\rangle \delta _{{\bf q,q}^{\prime }}\right] e^{i\omega t},
\label{28}
\end{eqnarray}
where 
\begin{eqnarray}
K({\bf q)} &=&\sum_{{\bf i}}K({\bf r}_{{\bf i}}-{\bf r}_{{\bf j}})e^{i{\bf %
q\cdot }({\bf r}_{{\bf i}}-{\bf r}_{{\bf j}})}  \nonumber \\
&=&K\sum_{i=1}^d\left( e^{iq_ia}+e^{-iq_ia}\right) =2K\sum_{i=1}^d\cos
(q_ia),  \label{29}
\end{eqnarray}
\begin{equation}
K(0{\bf )=}2Kd,  \label{30}
\end{equation}
and $a$ is the lattice constant, $q_i$ is the $i$th component of the wave
vector ${\bf q}$, and $d$ is the spatial dimensionality.

First we solve Eq. (\ref{27}). Obviously, it is a first-order linear
inhomogeneous differential equation with the canonical form 
\[
\frac{dy\left( t\right) }{dt}+P(t)y\left( t\right) =Q(t), 
\]
and its general solution is 
\[
y\left( t\right) =\frac 1{\mu (t)}\left[ \mu (t_0)y\left( t_0\right)
+\int_{t_0}^t\mu (\xi )Q(\xi )d\xi \right] , 
\]
where 
\[
\mu (t)=\exp \left[ \int P(t)dt\right] . 
\]
Applying it to Eq. (\ref{27}), one can get the following exact solution: 
\begin{eqnarray}
\left\langle M({\bf q}^{\prime },t)\right\rangle &=&\exp \left( -\frac t{%
\tau _{{\bf q}^{\prime }}}\right) \left[ \left\langle M({\bf q}^{\prime
},0)\right\rangle +\frac{NH_0\beta }b\delta _{{\bf q,q}^{\prime
}}\int_0^t\exp \left( \frac \xi {\tau _{{\bf q}^{\prime }}}+i\omega \xi
\right) d\xi \right]  \nonumber \\
&=&\left[ \left\langle M({\bf q}^{\prime },0)\right\rangle -H_0\chi (\omega ,%
{\bf q}^{\prime })\delta _{{\bf q,q}^{\prime }}\right] \exp \left( -\frac t{%
\tau _{{\bf q}^{\prime }}}\right) +H_0\chi (\omega ,{\bf q}^{\prime })\delta
_{{\bf q,q}^{\prime }}e^{i\omega t},  \label{31}
\end{eqnarray}
where 
\begin{equation}
\tau _{{\bf q}^{\prime }}=\frac 1{1-\left( 1/b\right) K({\bf q}^{\prime }%
{\bf )}}=\frac 1{1-\left( 2K/b\right) \sum_{i=1}^d\cos (q_i^{\prime }a)}
\label{32}
\end{equation}
and 
\begin{equation}
\chi (\omega ,{\bf q}^{\prime })=\frac{N\beta }b\cdot \frac 1{1-\left(
2K/b\right) \sum_{i=1}^d\cos (q_i^{\prime }a)+i\omega }  \label{33}
\end{equation}
are the wave-number-dependent relaxation time and the frequency- and
wave-number-dependent complex susceptibility, respectively.

From expression (\ref{32}) we can see that $\tau _{{\bf q}^{\prime }}$ is
finite for ${\bf q}^{\prime }\neq 0$ as the temperature approaches the
critical point $T_c$ $\left( K_c=J/k_\beta T_c=b/2d\right) $, while it
becomes to infinity for ${\bf q}^{\prime }=0$.

Substituting the solution (\ref{31}) of the local magnetization into Eq. (%
\ref{28}), the time evolution equation of the spin-pair correlation can be
rewritten as 
\begin{eqnarray}
&&\frac d{dt}\left\langle G({\bf q}^{\prime },{\bf q}^{\prime \prime
},t)\right\rangle +\left[ \left( 1-\frac 1bK({\bf q}^{\prime }{\bf )}\right)
+\left( 1-\frac 1bK({\bf q}^{\prime \prime }{\bf )}\right) \right]
\left\langle G({\bf q}^{\prime },{\bf q}^{\prime \prime },t)\right\rangle 
\nonumber \\
&=&\frac{NH_0\beta }b\left[ \left\langle M({\bf q}^{\prime },0)\right\rangle
-H_0\chi (\omega ,{\bf q}^{\prime })\delta _{{\bf q,q}^{\prime }}\right]
\delta _{{\bf q,q}^{\prime \prime }}\exp \left( -\frac t{\tau _{{\bf q}%
^{\prime }}}+i\omega t\right)  \nonumber \\
&&+\frac{NH_0\beta }b\left[ \left\langle M({\bf q}^{\prime \prime
},0)\right\rangle -H_0\chi (\omega ,{\bf q}^{\prime \prime })\delta _{{\bf %
q,q}^{\prime \prime }}\right] \delta _{{\bf q,q}^{\prime }}\exp \left( -%
\frac t{\tau _{{\bf q}^{\prime \prime }}}+i\omega t\right)  \nonumber \\
&&+\frac{NH_0\beta }b\left[ H_0\chi (\omega ,{\bf q}^{\prime })+H_0\chi
(\omega ,{\bf q}^{\prime \prime })\right] \delta _{{\bf q,q}^{\prime
}}\delta _{{\bf q,q}^{\prime \prime }}e^{2i\omega t}.  \label{34}
\end{eqnarray}
Equation (\ref{34}) is also a first-order linear inhomogeneous differential
equation. One can give its general solution 
\begin{eqnarray}
&&\left\langle G({\bf q}^{\prime },{\bf q}^{\prime \prime },t)\right\rangle 
\nonumber \\
&=&\exp \left( -\frac t{\tau _{{\bf q}^{\prime }}}-\frac t{\tau _{{\bf q}%
^{\prime \prime }}}\right) \left\{ \left\langle G({\bf q}^{\prime },{\bf q}%
^{\prime \prime },0)\right\rangle \right.  \nonumber \\
&&+\frac{NH_0\beta }b\left[ \left\langle M({\bf q}^{\prime },0)\right\rangle
-H_0\chi (\omega ,{\bf q}^{\prime })\delta _{{\bf q,q}^{\prime }}\right]
\delta _{{\bf q,q}^{\prime \prime }}\int_0^t\exp \left( \frac \xi {\tau _{%
{\bf q}^{\prime \prime }}}+i\omega \xi \right) d\xi  \nonumber \\
&&+\frac{NH_0\beta }b\left[ \left\langle M({\bf q}^{\prime \prime
},0)\right\rangle -H_0\chi (\omega ,{\bf q}^{\prime \prime })\delta _{{\bf %
q,q}^{\prime \prime }}\right] \delta _{{\bf q,q}^{\prime }}\int_0^t\exp
\left( \frac \xi {\tau _{{\bf q}^{\prime }}}+i\omega \xi \right) d\xi 
\nonumber \\
&&+\frac{NH_0\beta }b\left[ H_0\chi (\omega ,{\bf q}^{\prime })+H_0\chi
(\omega ,{\bf q}^{\prime \prime })\right] \delta _{{\bf q,q}^{\prime
}}\delta _{{\bf q,q}^{\prime \prime }}  \nonumber \\
&&\left. \times \int_0^t\exp \left( \frac \xi {\tau _{{\bf q}^{\prime }}}+%
\frac \xi {\tau _{{\bf q}^{\prime \prime }}}+2i\omega \xi \right) d\xi
\right\}  \nonumber \\
&=&\left[ \left\langle G({\bf q}^{\prime },{\bf q}^{\prime \prime
},0)\right\rangle -\left\langle M({\bf q}^{\prime },0)\right\rangle H_0\chi
(\omega ,{\bf q}^{\prime \prime })\delta _{{\bf q,q}^{\prime \prime
}}-\left\langle M({\bf q}^{\prime \prime },0)\right\rangle H_0\chi (\omega ,%
{\bf q}^{\prime })\delta _{{\bf q,q}^{\prime }}\right.  \nonumber \\
&&\left. +H_0^2\chi (\omega ,{\bf q}^{\prime })\chi (\omega ,{\bf q}^{\prime
\prime })\delta _{{\bf q,q}^{\prime }}\delta _{{\bf q,q}^{\prime \prime
}}\right] \exp \left( -\frac t{\tau _{{\bf q}^{\prime }}}-\frac t{\tau _{%
{\bf q}^{\prime \prime }}}\right)  \nonumber \\
&&+\left[ \left\langle M({\bf q}^{\prime },0)\right\rangle -H_0\chi (\omega ,%
{\bf q}^{\prime })\delta _{{\bf q,q}^{\prime }}\right] H_0\chi (\omega ,{\bf %
q}^{\prime \prime })\delta _{{\bf q,q}^{\prime \prime }}\exp \left( -\frac t{%
\tau _{{\bf q}^{\prime }}}+i\omega t\right)  \nonumber \\
&&+\left[ \left\langle M({\bf q}^{\prime \prime },0)\right\rangle -H_0\chi
(\omega ,{\bf q}^{\prime \prime })\delta _{{\bf q,q}^{\prime \prime
}}\right] H_0\chi (\omega ,{\bf q}^{\prime })\delta _{{\bf q,q}^{^{\prime
}}}\exp \left( -\frac t{\tau _{{\bf q}^{\prime \prime }}}+i\omega t\right) 
\nonumber \\
&&+H_0^2\chi (\omega ,{\bf q}^{\prime })\chi (\omega ,{\bf q}^{\prime \prime
})\delta _{{\bf q,q}^{\prime }}\delta _{{\bf q,q}^{\prime \prime
}}e^{2i\omega t}.  \label{35}
\end{eqnarray}

To make the solution an explicit one, we note that the factor $\exp \left(
-t/\tau _{{\bf q}}\right) $ can rewritten as 
\begin{eqnarray}
\exp \left( -\frac t{\tau _{{\bf q}}}\right) &=&\exp \left[ -\left( 1-\frac{%
2K}b\sum_{i=1}^d\cos (q_ia)\right) t\right]  \nonumber \\
&=&e^{-t}\prod_{i=1}^d\exp \left( \frac{2K}bt\frac{e^{iq_ia}+e^{-iq_ia}}2%
\right)  \nonumber
\end{eqnarray}
in which 
\[
\exp \left( \frac{2K}bt\frac{e^{iq_ia}+e^{-iq_ia}}2\right) 
\]
is just the generating function for the first-kind imaginary argument Bessel
function 
\begin{equation}
\exp \left[ \frac x2\left( \lambda +\lambda ^{-1}\right) \right]
=\sum_{\alpha =-\infty }^\infty \lambda ^\alpha I_\alpha \left( x\right) ,
\label{36}
\end{equation}
where $I_\alpha \left( x\right) $ is the first kind imaginary argument
Bessel function. Hence 
\begin{eqnarray}
\exp \left( -\frac t{\tau _{{\bf q}}}\right)
&=&e^{-t}\prod_{i=1}^d\sum_{n_i=-\infty }^\infty \left( e^{iq_ia}\right)
^{n_i}I_{n_i}\left( \frac{2K}bt\right)  \nonumber \\
&=&e^{-t}\prod_{i=1}^d\sum_{n_i=-\infty }^\infty
e^{iq_ix_{n_i}}I_{n_i}\left( \frac{2K}bt\right)  \nonumber \\
&=&e^{-t}\sum_{{\bf n}}e^{i{\bf q.r}_{{\bf n}}}I_{n_{_1}}\left( \frac{2K}b%
t\right) \cdots I_{n_{_d}}\left( \frac{2K}bt\right)  \label{37}
\end{eqnarray}
for convenience, where the summations for $n_{1,}\ldots ,n_d$ from $-\infty $
to $\infty $ are denoted by $\sum_{{\bf n}}$. Then Eqs. (\ref{31}) and (\ref
{35}) can be rewritten as 
\begin{eqnarray}
\left\langle M({\bf q}^{\prime },t)\right\rangle &=&\left[ \left\langle M(%
{\bf q}^{\prime },0)\right\rangle -H_0\chi (\omega ,{\bf q}^{\prime })\delta
_{{\bf q,q}^{\prime }}\right]  \nonumber \\
&&\times e^{-t}\sum_{{\bf n}}e^{i{\bf q.r}_{{\bf n}}}I_{n_{_1}}\left( \frac{%
2K}bt\right) \cdots I_{n_{_d}}\left( \frac{2K}bt\right) +H_0\chi (\omega ,%
{\bf q}^{\prime })\delta _{{\bf q,q}^{\prime }}e^{i\omega t},  \label{38}
\end{eqnarray}
\begin{eqnarray}
&&\left\langle G({\bf q}^{\prime },{\bf q}^{\prime \prime },t)\right\rangle 
\nonumber \\
&=&e^{-2t}\left[ \left\langle G({\bf q}^{\prime },{\bf q}^{\prime \prime
},0)\right\rangle -\left\langle M({\bf q}^{\prime },0)\right\rangle H_0\chi
(\omega ,{\bf q}^{\prime \prime })\delta _{{\bf q,q}^{\prime \prime
}}-\left\langle M({\bf q}^{\prime \prime },0)\right\rangle H_0\chi (\omega ,%
{\bf q}^{\prime })\delta _{{\bf q,q}^{\prime }}\right.  \nonumber \\
&&\left. +H_0^2\chi (\omega ,{\bf q}^{\prime })\chi (\omega ,{\bf q}^{\prime
\prime })\delta _{{\bf q,q}^{\prime }}\delta _{{\bf q,q}^{\prime \prime
}}\right]  \nonumber \\
&&\times \sum_{{\bf n,m}}e^{i{\bf q}^{\prime }{\bf .r}_{{\bf n}}+i{\bf q}%
^{\prime \prime }{\bf .r}_{{\bf m}}}I_{n_{_1}}\left( \frac{2K}bt\right)
I_{m_{_1}}\left( \frac{2K}bt\right) \cdots I_{n_{_d}}\left( \frac{2K}b%
t\right) I_{m_{_d}}\left( \frac{2K}bt\right)  \nonumber \\
&&+e^{-t}e^{i\omega t}\left[ \left\langle M({\bf q}^{\prime
},0)\right\rangle -H_0\chi (\omega ,{\bf q}^{\prime })\delta _{{\bf q,q}%
^{\prime }}\right] H_0\chi (\omega ,{\bf q}^{\prime \prime })\delta _{{\bf %
q,q}^{\prime \prime }}  \nonumber \\
&&\times \sum_{{\bf n}}e^{i{\bf q}^{\prime }{\bf .r}_{{\bf n}%
}}I_{n_{_1}}\left( \frac{2K}bt\right) \cdots I_{n_{_d}}\left( \frac{2K}b%
t\right)  \nonumber \\
&&+e^{-t}e^{i\omega t}\left[ \left\langle M({\bf q}^{\prime \prime
},0)\right\rangle -H_0\chi (\omega ,{\bf q}^{\prime \prime })\delta _{{\bf %
q,q}^{\prime \prime }}\right] H_0\chi (\omega ,{\bf q}^{\prime })\delta _{%
{\bf q,q}^{^{\prime }}}  \nonumber \\
&&\times \sum_{{\bf n}}e^{i{\bf q}^{\prime \prime }{\bf .r}_{{\bf n}%
}}I_{n_{_1}}\left( \frac{2K}bt\right) \cdots I_{n_{_d}}\left( \frac{2K}b%
t\right)  \nonumber \\
&&+H_0^2\chi (\omega ,{\bf q}^{\prime })\chi (\omega ,{\bf q}^{\prime \prime
})\delta _{{\bf q,q}^{\prime }}\delta _{{\bf q,q}^{\prime \prime
}}e^{2i\omega t}.  \label{39}
\end{eqnarray}
Taking the inverse Fourier transformation 
\begin{equation}
\left\langle \sigma _{{\bf k}}\left( t\right) \right\rangle =\frac 1N\sum_{%
{\bf q}^{\prime }}\left\langle M({\bf q}^{\prime },t)\right\rangle e^{-i{\bf %
q}^{\prime }{\bf \cdot r}_{{\bf k}}},  \label{40}
\end{equation}
\begin{equation}
\left\langle \sigma _{{\bf k}}\left( t\right) \sigma _{{\bf l}}\left(
t\right) \right\rangle =\frac 1{N^2}\sum_{{\bf q}^{\prime },{\bf q}^{\prime
\prime }}\left\langle G({\bf q}^{\prime },{\bf q}^{\prime \prime
},t)\right\rangle e^{-i{\bf q}^{\prime }{\bf \cdot r}_{{\bf k}}-i{\bf q}%
^{\prime \prime }{\bf \cdot r}_{{\bf l}}},  \label{41}
\end{equation}
and using the following relation: 
\begin{equation}
\frac 1N\sum_{{\bf q}}e^{i{\bf q\cdot }({\bf r}_{{\bf i}}-{\bf r}_{{\bf j}%
})}=\delta _{{\bf i,j}},  \label{42}
\end{equation}
the local magnetization and the pair correlation of the $d$-dimensional
hypercubic system can be written as 
\begin{eqnarray}
\left\langle \sigma _{{\bf k}}(t)\right\rangle &=&e^{-t}\sum_{{\bf n}%
}\left\{ \left\langle \sigma _{{\bf n}}(0)\right\rangle I_{k_1-n_1}\left( 
\frac{2K}bt\right) \cdots I_{k_d-n_d}\left( \frac{2K}bt\right) \right. 
\nonumber \\
&&-\left. \frac 1NH_0\chi (\omega ,{\bf q})e^{i{\bf q\cdot r}_{{\bf n}%
}}I_{k_1+n_1}\left( \frac{2K}bt\right) \cdots I_{k_d+n_d}\left( \frac{2K}b%
t\right) \right\}  \nonumber \\
&&+\frac 1NH_0\chi (\omega ,{\bf q})e^{i\omega t-i{\bf q\cdot r}_{{\bf k}}},
\label{43}
\end{eqnarray}
and 
\begin{eqnarray}
&&\left\langle \sigma _{{\bf k}}\left( t\right) \sigma _{{\bf l}}\left(
t\right) \right\rangle  \nonumber \\
&=&e^{-2t}\sum_{{\bf n,m}}\left\langle \sigma _{{\bf n}}\left( 0\right)
\sigma _{{\bf m}}\left( 0\right) \right\rangle I_{k_1-n_{_1}}\left( \frac{2K}%
bt\right) I_{l_1-m_{_1}}\left( \frac{2K}bt\right) \cdots I_{k_d-n_d}\left( 
\frac{2K}bt\right) I_{l_d-m_d}\left( \frac{2K}bt\right)  \nonumber \\
&&-\frac 1Ne^{-2t}H_0\chi (\omega ,{\bf q})e^{-i{\bf q\cdot r}_{{\bf l}%
}}\sum_{{\bf n,m}}\left\langle \sigma _{{\bf n}}\left( 0\right)
\right\rangle e^{i{\bf q}\cdot {\bf r}_{{\bf m}}}  \nonumber \\
&&\times I_{k_1-n_{_1}}\left( \frac{2K}bt\right) I_{m_{_1}}\left( \frac{2K}b%
t\right) \cdots I_{k_d-n_d}\left( \frac{2K}bt\right) I_{m_{_d}}\left( \frac{%
2K}bt\right)  \nonumber \\
&&-\frac 1Ne^{-2t}H_0\chi (\omega ,{\bf q})e^{-i{\bf q\cdot r}_{{\bf k}%
}}\sum_{{\bf n,m}}\left\langle \sigma _{{\bf m}}\left( 0\right)
\right\rangle e^{i{\bf q}\cdot {\bf r}_{{\bf n}}}  \nonumber \\
&&\times I_{n_{_1}}\left( \frac{2K}bt\right) I_{l_1-m_{_1}}\left( \frac{2K}b%
t\right) \cdots I_{n_{_d}}\left( \frac{2K}bt\right) I_{l_d-m_{_d}}\left( 
\frac{2K}bt\right)  \nonumber \\
&&+\frac 1{N^2}e^{-2t}H_0^2\chi ^2(\omega ,{\bf q})\sum_{{\bf n,m}}e^{i{\bf q%
}\cdot {\bf r}_{{\bf n+m}}}  \nonumber \\
&&\times I_{k_1+n_{_1}}\left( \frac{2K}bt\right) I_{l_1+m_{_1}}\left( \frac{%
2K}bt\right) \cdots I_{k_d+n_{_d}}\left( \frac{2K}bt\right)
I_{l_d+m_d}\left( \frac{2K}bt\right)  \nonumber \\
&&+\frac 2Ne^{-t}e^{i\omega t}H_0\chi (\omega ,{\bf q})e^{-i{\bf q\cdot r}_{%
{\bf l}}}\sum_{{\bf n}}\left\langle \sigma _{{\bf n}}\left( 0\right)
\right\rangle I_{k_1-n_{_1}}\left( \frac{2K}bt\right) \cdots
I_{k_d-n_{_d}}\left( \frac{2K}bt\right)  \nonumber \\
&&-\frac 2{N^2}e^{-t}e^{i\omega t}H_0^2\chi ^2(\omega ,{\bf q})e^{-i{\bf %
q\cdot r}_{{\bf l}}}\sum_{{\bf n}}e^{i{\bf q\cdot r}_{{\bf n}%
}}I_{k_1+n_{_1}}\left( \frac{2K}bt\right) \cdots I_{k_d+n_{_d}}\left( \frac{%
2K}bt\right)  \nonumber \\
&&+\frac 1{N^2}H_0^2\chi ^2(\omega ,{\bf q})e^{2i\omega t-i{\bf q\cdot r}_{%
{\bf k}+{\bf l}}},  \label{44}
\end{eqnarray}
respectively. Because $n_i$ (or $m_i$) can take any real value in the region 
$\left( -\infty ,\infty \right) $, the summations for $n_i$ and $-$ $n_i$
(or for $m_i$ and $-m_i$) are equal. In addition, the summation indexes $%
{\bf n}$ and ${\bf m}$ can exchange each other. Then, the Eqs. (\ref{43})
and (\ref{44}) can be rewritten as 
\begin{eqnarray}
\left\langle \sigma _{{\bf k}}(t)\right\rangle &=&e^{-t}\sum_{{\bf n}}\left[
\left\langle \sigma _{{\bf n}}(0)\right\rangle -\frac 1NH_0\chi (\omega ,%
{\bf q})e^{i{\bf q\cdot r}_{{\bf n}}}\right] I_{k_1-n_1}\left( \frac{2K}b%
t\right) \cdots I_{k_d-n_d}\left( \frac{2K}bt\right)  \nonumber \\
&&+\frac 1NH_0\chi (\omega ,{\bf q})e^{i\omega t-i{\bf q\cdot r}_{{\bf k}}},
\label{45}
\end{eqnarray}
\begin{eqnarray}
\left\langle \sigma _{{\bf k}}\left( t\right) \sigma _{{\bf l}}\left(
t\right) \right\rangle &=&e^{-2t}\sum_{{\bf n,m}}\left[ \left\langle \sigma
_{{\bf n}}\left( 0\right) \sigma _{{\bf m}}\left( 0\right) \right\rangle +%
\frac 1{N^2}H_0^2\chi ^2(\omega ,{\bf q})e^{-i{\bf q}\cdot {\bf r}_{{\bf n-m}%
}}\right]  \nonumber \\
&&\times I_{k_1-n_{_1}}\left( \frac{2K}bt\right) I_{l_1-m_{_1}}\left( \frac{%
2K}bt\right) \cdots I_{k_d-n_d}\left( \frac{2K}bt\right) I_{l_d-m_d}\left( 
\frac{2K}bt\right)  \nonumber \\
&&-\frac 1Ne^{-2t}H_0\chi (\omega ,{\bf q})\left[ e^{-i{\bf q\cdot r}_{{\bf k%
}}}+e^{-i{\bf q\cdot r}_{{\bf l}}}\right]  \nonumber \\
&&\times \sum_{{\bf n,m}}\left\langle \sigma _{{\bf n}}\left( 0\right)
\right\rangle e^{i{\bf q}\cdot {\bf r}_{{\bf m}}}I_{k_1-n_{_1}}\left( \frac{%
2K}bt\right) I_{m_{_1}}\left( \frac{2K}bt\right) \cdots I_{k_d-n_d}\left( 
\frac{2K}bt\right) I_{m_{_d}}\left( \frac{2K}bt\right)  \nonumber \\
&&+\frac 2Ne^{-t}e^{i\omega t}H_0\chi (\omega ,{\bf q})e^{-i{\bf q\cdot r}_{%
{\bf l}}}\sum_{{\bf n}}\left[ \left\langle \sigma _{{\bf n}}\left( 0\right)
\right\rangle -H_0\chi (\omega ,{\bf q})e^{-i{\bf q\cdot r}_{{\bf n}}}\right]
\nonumber \\
&&\times I_{k_1-n_{_1}}\left( \frac{2K}bt\right) \cdots I_{k_d-n_{_d}}\left( 
\frac{2K}bt\right)  \nonumber \\
&&+\frac 1{N^2}H_0^2\chi ^2(\omega ,{\bf q})e^{2i\omega t-i{\bf q\cdot r}_{%
{\bf k}+{\bf l}}},  \label{46}
\end{eqnarray}
where $\left\langle \sigma _{{\bf n}}\left( 0\right) \right\rangle $ and $%
\left\langle \sigma _{{\bf n}}\left( 0\right) \sigma _{{\bf m}}\left(
0\right) \right\rangle $ correspond to their initial values.

So far we have obtained the exact solutions of the $d$-dimensional kinetic
Gaussian model in the periodic external field. Letting $H_0=0$, one can get

\begin{equation}
\left\langle \sigma _{k_1\ldots k_d}(t)\right\rangle
=e^{-t}\sum_{n_{1,}\ldots ,n_d=-\infty }^\infty \left\langle \sigma
_{n_1\ldots n_d}(0)\right\rangle I_{k_1-n_1}\left( \frac{2K}bt\right) \cdots
I_{k_d-n_d}\left( \frac{2K}bt\right) ,  \label{47}
\end{equation}
\begin{eqnarray}
\left\langle \sigma _{k_1\ldots k_d}(t)\sigma _{l_1\ldots
l_d}(t)\right\rangle &=&e^{-2t}\sum_{n_{1,}\ldots ,n_d=-\infty }^\infty
\sum_{m_{1,}\ldots ,m_d=-\infty }^\infty \left\langle \sigma _{n_1\ldots
n_d}(0)\sigma _{m_1\ldots m_d}(0)\right\rangle  \nonumber \\
&&\times I_{k_1-n_{_1}}\left( \frac{2K}bt\right) I_{l_1-m_{_1}}\left( \frac{%
2K}bt\right) \cdots I_{k_d-n_d}\left( \frac{2K}bt\right) I_{l_d-m_d}\left( 
\frac{2K}bt\right) .  \label{48}
\end{eqnarray}
Equations (\ref{47}) and (\ref{48}) are just the exact solutions of the $d$%
-dimensional kinetic Gaussian model in zero external field, which agree with
the results obtained in our previous paper\cite{Zhu}.

To understand the physical meaning of the Eqs. (\ref{47}) and (\ref{48}), it
is interesting to investigate the long-time asymptotic behavior of the local
magnetization and the spin-pair-correlation function. By use of the
asymptotic expansion expression of the first kind of imaginary argument
Bessel function 
\begin{equation}
I_v(x)=\frac{e^x}{\sqrt{2\pi x}}\sum_{n=0}^\infty \frac{(-)^n(v,n)}{(2x)^n}+%
\frac{e^{-x+(v+\frac 12)\pi i}}{\sqrt{2\pi x}}\sum_{n=0}^\infty \frac{(v,n)}{%
(2x)^n},(-\pi /2<\arg x<3\pi /2),\left| x\right| \rightarrow \infty ,
\label{49}
\end{equation}
where 
\[
(v,n)=\frac{\Gamma \left( \frac 12+v+n\right) }{n!\Gamma \left( \frac 12%
+v-n\right) }, 
\]
one can get 
\begin{equation}
\left\langle \sigma _{{\bf k}}(t)\right\rangle \sim \sum_{{\bf n}}\left[
\left\langle \sigma _{{\bf n}}(0)\right\rangle -\frac 1NH_0\chi (\omega ,%
{\bf q})e^{i{\bf q\cdot r}_{{\bf n}}}\right] \frac 1{t^{d/2}}e^{-t/\tau }+%
\frac 1NH_0\chi (\omega ,{\bf q})e^{i\omega t-i{\bf q\cdot r}_{{\bf k}}},
\label{50}
\end{equation}
\begin{eqnarray}
\left\langle \sigma _{{\bf k}}\left( t\right) \sigma _{{\bf l}}\left(
t\right) \right\rangle &\sim &\sum_{{\bf n,m}}\left[ \left\langle \sigma _{%
{\bf n}}\left( 0\right) \sigma _{{\bf m}}\left( 0\right) \right\rangle +%
\frac 1{N^2}H_0^2\chi ^2(\omega ,{\bf q})e^{-i{\bf q}\cdot {\bf r}_{{\bf n-m}%
}}\right.  \nonumber \\
&&\left. -\frac 1NH_0\chi (\omega ,{\bf q})\left( e^{-i{\bf q\cdot r}_{{\bf %
k-m}}}+e^{-i{\bf q\cdot r}_{{\bf l-m}}}\right) \left\langle \sigma _{{\bf n}%
}\left( 0\right) \right\rangle \right] \frac 1{t^d}e^{-2t/\tau }  \nonumber
\\
&&+\frac 2NH_0\chi (\omega ,{\bf q})e^{i\omega t-i{\bf q\cdot r}_{{\bf l}%
}}\sum_{{\bf n}}\left[ \left\langle \sigma _{{\bf n}}\left( 0\right)
\right\rangle -H_0\chi (\omega ,{\bf q})e^{-i{\bf q\cdot r}_{{\bf n}%
}}\right] \frac 1{t^{d/2}}e^{-t/\tau }  \nonumber \\
&&+\frac 1{N^2}H_0^2\chi ^2(\omega ,{\bf q})e^{2i\omega t-i{\bf q\cdot r}_{%
{\bf k}+{\bf l}}},  \label{51}
\end{eqnarray}
where 
\begin{equation}
\tau =\frac 1{1-2kd/b},  \label{52}
\end{equation}
is the relaxation time of the system. From Eqs. (\ref{50}) and (\ref{51}) we
can see that both the local magnetization and the spin-pair-correlation
function consist of two parts: one decays with $t$, and the other vibrates
with $t$. Since $\tau $ increases to infinity as the temperature approaches
the static critical point $T_c$ $\left( K_c=J/k_\beta T_c=b/2d\right) ,$ the
decay term will occur critical slowing down phenomenon.

We now turn on the response of the system to the time-dependent external
field. According to the general theory of linear response, the complex
susceptibility $\chi (\omega ,{\bf q})$ is expressed in terms of the
equilibrium correlation of magnetization, namely\cite{Kubo} 
\begin{equation}
\chi (\omega ,{\bf q})=\chi (0,{\bf q})-\frac{i\omega }{k_\beta T}%
\int_0^\infty \left\langle M(-{\bf q},0)M({\bf q},t)\right\rangle
_ee^{-i\omega t}dt,  \label{53}
\end{equation}
where 
\begin{equation}
\chi (0,{\bf q})=\frac 1{k_\beta T}\left\langle M(-{\bf q},0)M({\bf q}%
,0)\right\rangle _e,  \label{54}
\end{equation}
and $\left\langle \cdots \right\rangle _e$ denotes the average over
equilibrium distribution.

Because 
\begin{equation}
\chi (\omega ,{\bf q})=\frac{N\beta }b\cdot \frac 1{1-\frac{2K}b%
\sum_{i=1}^d\cos (q_ia)+i\omega },  \label{55}
\end{equation}
Eqs. (\ref{53}) and (\ref{54}), therefore, mean that 
\begin{equation}
\left\langle M(-{\bf q},0)M({\bf q},t)\right\rangle _e=\left\langle M(-{\bf q%
},0)M({\bf q},0)\right\rangle _e\exp \left( -\frac t{\tau _{{\bf q}}}\right)
,  \label{56}
\end{equation}
and 
\begin{equation}
\left\langle M(-{\bf q},0)M({\bf q},0)\right\rangle _e=\frac N{b\left[ 1-%
\frac 1bK({\bf q)}\right] }=\frac N{b\left[ 1-\frac{2K}b\sum_{i=1}^d\cos
(q_ia)\right] },  \label{57}
\end{equation}
where 
\begin{equation}
\tau _{{\bf q}}=\frac 1{1-\frac 1bK({\bf q)}}=\frac 1{1-\frac{2K}b%
\sum_{i=1}^d\cos (q_ia)}.  \label{58}
\end{equation}
It is interesting to note that as the temperature approaches the static
critical point $T_c$, for ${\bf q}=0$ the static spatial correlation
diverges, while for ${\bf q}\neq 0$ it remains finite. However, whether the
singularity occurs or not, the dynamic responses strongly depend on the
frequency $\omega $ and the wave vector ${\bf q}$ of the external field.

\section{Concluding remarks}

\label{Sec.4}In this paper, the single-spin transition Glauber dynamics is
used to investigate the kinetic Gaussian model in a periodic external field.
We have exactly obtained the local magnetization and equal-time pair
correlation function of the $d$-dimensional isotropic hypercubic lattice
Gaussian model by using Fourier's transformation. The related critical
dynamics characteristics of the system are discussed.

The master equation (\ref{1}) with the transition probability given by Eq. (%
\ref{8}) conserves the important features of a cooperative system. When the
system is in a time-dependent external field, the dynamical model itself
cannot be exact at very high-frequencies, it is only suitable for the case
of low frequencies. In fact, the existence of a high frequency field weakens
stochastic motion and makes a thermal equilibrium state with canonical
distribution impossible. Even so, it would not bring any impact for the
characteristic behavior of the system at low frequencies. The present work
has clearly shown that the local magnetization, the equal-time pair
correlation, and the dynamical responses of the system to a time-dependent
external field strongly depend on the frequency and the wave vector of the
external field, and when $\omega \rightarrow 0$ and ${\bf q\rightarrow 0}$,
they approach static results.

The Gaussian model is certainly an idealization, but it is interesting and
simple enough to obtain some fundamental knowledge of dynamical process in
cooperative systems. Furthermore, although it is an extension of Ising
model, the Gaussian model is quite different from Ising model in the
properties of the phase transition\cite{Li}. It is well known that in the
equilibrium case the Gaussian model is exactly solvable on translational
invariant lattices\cite{Stanley}. Meantime, as we have done, the Gaussian
model is also exactly solvable in dynamical case\cite{Zhu}. Finally, we can
anticipate that the kinetic Gaussian model will be a starting point to study
the kinetic $s^4$ (or $\phi ^4$ ) model.

\section{Acknowledgments}

This work is supported by the National Basic Research Project ``Nonlinear
Science'' and the National Natural Science Foundation of China under Grant
No. 19775008. J.Y.Z. thanks Dr. Z. Gao for his valuable discussions.

\end{document}